\documentclass[letterpaper, 10 pt, conference]{ieeetran}  

\IEEEoverridecommandlockouts                              

\usepackage{epsfig} 
\usepackage{varwidth}
\usepackage[abs]{overpic}
\usepackage{algorithmic}

\usepackage[caption=false,font=footnotesize]{subfig}
\newcounter{numberlatin}
\newcommand{\rom}[1]{\romannumeral #1}
\usepackage{cite}

\usepackage{enumerate}

\usepackage{epstopdf}
\usepackage{color}

\usepackage{amsmath}
\usepackage{amssymb}  

\usepackage{algorithmic}
\usepackage{caption}

\usepackage[caption=false,font=footnotesize]{subfig}
\begin{document}

%
\title{\LARGE \bf On executing aggressive quadrotor attitude tracking maneuvers\\ under actuator constraints
 }
%
%
%

\author{Michalis~Ramp$^1$~and~Evangelos~Papadopoulos$^2$,~\IEEEmembership{Senior~Member,~IEEE}
\thanks{$^1$M. Ramp is with the Department of Mechanical Engineering, National Technical University of Athens, (NTUA) 15780 Athens, Greece.
{\tt\small rampmich@mail.ntua.gr}}
\thanks{$^2$E. Papadopoulos is with the Department of Mechanical Engineering, NTUA, 15780 Athens (tel: +30-210-772-1440; fax: +30-210-772-1455).
{\tt\small egpapado@central.ntua.gr}}
}

\IEEEpubid{Accepted for publication in the Mediteranean Conference on Control and Automation (MED18)}
\IEEEpubidadjcol


\maketitle

\begin{abstract}

The  quadrotor task of negotiating aggressive attitude maneuvers while adhering to motor constraints is addressed here.
The majority of high level quadrotor Nonlinear Control System (NCS) solutions ignore motor control authority limitations, especially important during aggressive attitude maneuvers, generating unrealizable thrusts and negating the validity of the accompanying stability proofs.
Here, an attitude control framework is developed, comprised by a thrust allocation strategy and a specially designed geometric attitude tracking controller, allowing the quadrotor to achieve aggressive attitude maneuvers, while complying to actuator constraints and simultaneously staying ''close'' to a desired position command in a computationally inexpensive way.
This is a novel contribution resulting in thrusts realizable by available quadrotors during aggressive attitude maneuvers, and enhanced performance guaranteed by valid stability proofs.
Also, it is shown that the developed controller can be combined with a collective thrust expression in producing a position/yaw tracking controller.
Through rigorous stability proofs, both the position and attitude frameworks are shown to have desirable closed loop properties that are almost global.
This establishes a quadrotor control solution allowing the vehicle to negotiate aggressive maneuvers position/attitude on SE(3).
Simulations illustrate and validate the effectiveness and capabilities of the developed solution.
\end{abstract}


%
\IEEEpeerreviewmaketitle

\section{Introduction}
%
%
%
%
Quadrotor unmanned aerial vehicles (UAV), composed of two pairs of counter rotating outrunner motor/propeller assemblies, produced a low cost and agile vertical takeoff and landing (VTOL) platform characterized by high thrust to weight ratio, suitable to negotiate an extensive range of flight scenarios and applications.
Since a quadrotor has only four inputs, it is underactuated;
it can track at most four degrees of freedom (dof) although it has six.

The flight control of quadrotors is carried out, (a) first through a high level control solution that produces the needed collective thrust and torque control efforts, followed by (b) their resolution into the single rotor thrusts by means of \textit{thrust allocation/mapping}, and (c) the conversion of the single rotor thrusts to PWM signals that are fed to Electronic Speed Controlers (ESC) that drive the motors.

A plethora of high level quadrotor controllers including (but not limited to) backsteping \cite{MahonyHamel}, geometric \cite{geomquadlee}, \cite{geomquadlee_asian}, \cite{qeopidfar} and hybrid global/robust controllers \cite{Casau}, \cite{Abdessameud}, \cite{NaldiRob} have been developed; however the task of thrust allocation/mapping has received little attention, despite the fact that the flight performance of a multi-rotor vehicle is interconnected with the control allocation strategy.
Specifically the control allocation problem as means of avoiding motor saturation has not been studied in depth;
motor limitations are usually tackled using global constrained optimization through the enforcement of actuator constraints between waypoints \cite{Mellinger}, \cite{gloia}.
A few works treat the control allocation problem as means of avoiding motor saturation \cite{Monteiro}, \cite{Faessler}, \cite{Zaki}.
To handle infeasible inputs, a control allocation method prioritizing the generated body torques over collective thrust was developed \cite{Monteiro}.
A saturation scheme prioritizing control inputs according to their importance in regards to trajectory tracking was presented and validated experimentally \cite{Faessler}.
A control allocation-like scheme utilizing nonlinear constraint optimization at each control iteration, extracts reference angles for the desired trajectory, based on generalized commands, was proposed and validated in simulation \cite{Zaki}.

The source of inspiration for this work was the study of the simulation results from high level geometric controllers \cite{geomquadlee}, \cite{geomquadlee_asian}, \cite{qeopidfar}.
These controllers achieve aggressive quadrotor maneuvers through the concatenated use of two flight modes; a \textit{Position Mode} able to track a desired CM/yaw trajectory, and an \textit{Attitude Mode} used for short durations of time, to track a desired quadrotor attitude.
Studying the impressive results from the aforementioned publications, it was noticed that thrust saturation rarely occurs during the Position Mode, provided that the position maneuver is smooth and of reasonable rate.
In contrast, during the Attitude Mode, if large angle attitude tracking is the desired task, the motors saturate even if the rate of the maneuver is very slow.

This observation led us to develop an attitude allocation strategy and a tracking controller to be used during the Attitude Mode, allowing the quadrotor to achieve (a) precise attitude tracking and simultaneously (b) comply to actuator constraints while (c) staying ''close'' to a desired position command in a computationally inexpensive manner.
In addition, the supplemented attitude stability proofs from \cite{geomquadlee}, \cite{geomquadlee_asian}, \cite{qeopidfar}, do not account for motor saturations;
thus in order for the stability assurances (regions of attraction) to be valid, the demanded control effort must be available, i.e. saturation must not take place.
Using our developed allocation strategy and controller, this limitation is bypassed (thus the stability assurances are valid), allowing the commanded thrusts during attitude maneuvers to be realizable by the majority of quadrotors produced in the industry, introducing an important contribution.
\IEEEpubidadjcol
The proposed strategy and controller are validated in simulation in the presence of motor saturations.

The paper is structured as follows.
Section \ref{sec:kinetics} outlines the dynamic model.
Section \ref{overA} describes the design of the attitude allocation strategy/tracking controller.
Section \ref{conmodpos} details a complete high level quadrotor locomotion scheme, obtained through the development of a position ontroller that employs the attitude controller of Sec. \ref{overA}. 
Section \ref{simulation} shows results validating the controllers, strategy and claims from Sec. \ref{overA}.
Concluding remarks wrap up the paper.

\section{Quadrotor Kinetics Model\label{sec:kinetics}}
\begin{figure}[h!]
\centering
\includegraphics[width=1\columnwidth]{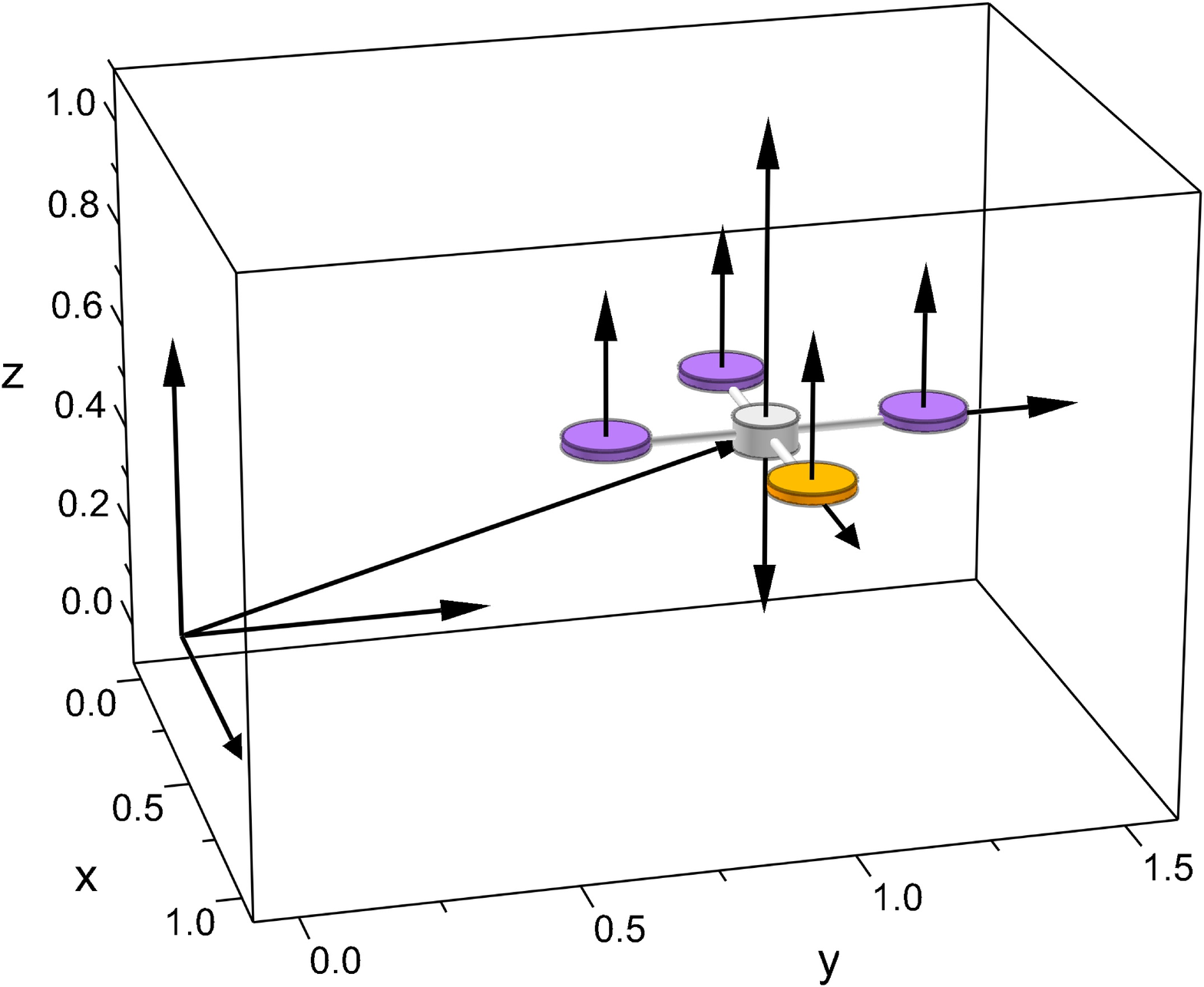}
\put(-115,57){\parbox{0.4\linewidth}{$-mg\mathbf{E}_{3}$}}
\put(-85,66){\parbox{\columnwidth}{$\mathbf{e}_{1}$}}
\put(-48,93){\parbox{0.4\linewidth}{$\mathbf{e}_{2}$}}
\put(-101,140){\parbox{0.4\linewidth}{$\mathbf{e}_{3}$}}
\put(-150,77){\parbox{0.4\linewidth}{$\mathbf{x}$}}
\put(-180,32){\parbox{\columnwidth}{$\mathbf{E}_{1}$}}
\put(-140,62){\parbox{0.4\linewidth}{$\mathbf{E}_{2}$}}
\put(-194,107){\parbox{0.4\linewidth}{$\mathbf{E}_{3}$}}
\put(-94,111){\parbox{0.4\linewidth}{$f_{1}$}}
\put(-78,121){\parbox{0.4\linewidth}{$f_{2}$}}
\put(-109,126){\parbox{0.4\linewidth}{$f_{3}$}}
\put(-128,116){\parbox{0.4\linewidth}{$f_{4}$}}
\caption{Quadrotor with its coordinate frames, motor thrusts, and weight.}
\label{Quadrotor}
\end{figure}

A quadrotor system utilizes two pairs of counter rotating out-runner motor/propeller assemblies, see Fig. \ref{Quadrotor}.
Thrust and torque are generated normal to the plane produced by the centers of mass (CM) of the rotors. 
A body-fixed frame I$_{b}\big\{\mathbf{e}_1,\mathbf{e}_2,\mathbf{e}_3\big\}$ and an inertial frame of reference I$_{R}\big\{\mathbf{E}_1,\mathbf{E}_2,\mathbf{E}_3\big\}$ are used.
The origin of I$_{b}$ is located at the quadrotor CM and its first two axes, $\mathbf{e}_1$, $\mathbf{e}_2$, are parallel with two quadrotor legs, see Fig. \ref{Quadrotor}, lying on the same plane defined by the CM of the rotors and the quadrotor CM.

The following assumptions hold throughout the paper.
It is assumed that the actual control input is the thrust of each propeller, which is parallel to the $\mathbf{e}_{3}$, acting on the extremity of each quadrotor leg.
Propellers 1, 3, produce positive thrust along $\mathbf{e}_{3}$ by rotating clockwise, and propellers 2, 4, produce positive thrust along the direction of $\mathbf{e}_{3}$ when rotating counterclockwise.
The collective thrust is denoted by $f=\sum_{i=1}^{4} f_i\in\mathbb{R}$ and is positive along $\mathbf{e}_{3}$ ($f_{i}$ and other system variables are defined in Table \ref{Table}).

\begin{table}[h!]
\caption{Definitions of variables.}
\label{Table}
\begin{center}
\begin{tabular}{|l|l|}
\hline
$\mathbf{x}\in\mathbb{R}^3$ & Position of quadrotor CM  wrt. I$_R$ in I$_R$\\
\hline
$\mathbf{v}\in\mathbb{R}^3$ & Velocity of quadrotor CM wrt. I$_R$ in I$_R$\\
\hline
$^{b}\boldsymbol{\omega}\in\mathbb{R}^3$ & Angular velocity of the quadrotor wrt. I$_R$ in I$_{b}$\\
\hline
$\mathbf{R}\in\text{SO}\left(3\right)$ & Rotation matrix from $\mathbf{I}_b$ to $\mathbf{I}_R$ frame\\
\hline
$^b\mathbf{u}\in\mathbb{R}^3$ & Control moment $^b\mathbf{u}{=}[{}^{b}u_{1};{}^{b}u_{2};{}^{b}u_{3}]$ in I$_b$\\
\hline
$f_i\in\mathbb{R}$ & $i^{th}$ propeller thrust along $\mathbf{e}_{3}$ \\
\hline
$b_T\in\mathbb{R}^{+}$ & Torque coefficient \\
\hline
$g\in\mathbb{R}$ & Gravity constant\\
\hline
$d\in\mathbb{R}^{+}$ & Distance between each motor CM and quadrotor CM\\
\hline
$\mathbf{J}\in\mathbb{R}^{3\times3}$ & Inertial matrix (IM) of the quadrotor in I$_b$\\
\hline
$m\in\mathbb{R}$ & System total mass\\
\hline
$\lambda_{min,max}(.)$ & Minimum, maximum eigenvalue of $(.)$ respectively\\
\hline
\end{tabular}
\end{center}
\end{table}

The $i^{th}$ motor torque, $\boldsymbol{\tau}_{i}$, due to its propeller, is assumed to be proportional to the thrust $f_{i}$ and is given by, \cite{Tayebi},
\begin{IEEEeqnarray*}{C}
\boldsymbol{\tau}_{i}=(-1)^{i}b_{T}f_{i}\mathbf{e}_{3}, i=1,..,4
\end{IEEEeqnarray*}
The collective thrust $f$ and moment vector, $^{b}\mathbf{u}$, are given by,
\begin{IEEEeqnarray}{C}
\begin{bmatrix}
f\\
^b\mathbf{u}
\end{bmatrix}
=
\begin{bmatrix}
1&1&1&1\\
0&d&0&-d\\
-d&0&d&0\\
-b_{T}&b_{T}&-b_{T}&b_{T}\\
\end{bmatrix}\!\!\mathbf{F};\;
\mathbf{F}=\begin{bmatrix}
f_{1}\\
f_{2}\\
f_{3}\\
f_{4}\\
\end{bmatrix}\IEEEeqnarraynumspace
\label{eq:distribution}
\end{IEEEeqnarray}
where $\mathbf{F}\in\mathbb{R}^{4}$ is the thrust vector, and the $4\times4$ matrix is always of full rank for $d,b_{T}\in\mathbb{R}^{+}$.
The collective thrust $f$ and torque $^{b}\mathbf{u}$ will be regarded as control inputs and the thrust for each propeller will be calculated using (\ref{eq:distribution}).

The quadrotor configuration is described by its attitude wrt. $\mathbf{I}_{R}$, and the position vector of its CM, again wrt. $\mathbf{I}_{R}$.
The configuration manifold is the special Euclidean group SE(3)=$\mathbb{R}^{3}\times\text{SO(3)}$.
The collective thrust vector produced by the propellers, in $\mathbf{I}_{R}$, is given by $\mathbf{R}f\mathbf{e}_{3}$.
The equations of motion of the system are given by,
\begin{IEEEeqnarray}{rCl}
\dot{\mathbf{x}}&=&{}\mathbf{v}\IEEEnonumber\IEEEeqnarraynumspace\\
m\dot{\mathbf{v}}&=&-mg\mathbf{E}_{3}+\mathbf{R}f\mathbf{e}_{3}\IEEEyesnumber\label{eq:position}\\
\mathbf{J}{}^{b}\dot{\boldsymbol{\omega}}&=&{}^{b}\mathbf{u}-{}^{b}\boldsymbol{\omega}\times\mathbf{J}{}^{b}\boldsymbol{\omega}\IEEEyesnumber\IEEEeqnarraynumspace\label{eq:attitude}\\
\dot{\mathbf{R}}&=&\mathbf{R}S({}^{b}\boldsymbol{\omega}) \IEEEyesnumber
\label{attitude_kinem}\end{IEEEeqnarray}
and $S(.):\mathbb{R}^{3}{\rightarrow}\mathfrak{so}(3)$ is the cross product map described by,
\begin{IEEEeqnarray}{rCl}
\begin{array}{c}
S(\mathbf{r}){=}[{0},{-r_{3}},{r_{2}};{r_{3}},{0},{-r_{1}};{-r_{2}},{r_{1}},0]\\
{S^{-1}}(S(\mathbf{r})){=}\mathbf{r}
\end{array}\label{iso}
\end{IEEEeqnarray}

\section{Exploiting Attitude Dynamics Redundancy\label{overA}} 

As mentioned earlier, during aggressive attitude maneuvers, existing geometric controllers and other Nonlinear Control System (NCS) solutions produce negative thrusts (see \cite{geomquadlee}, \cite{geomquadlee_asian}, \cite{qeopidfar}) that are not realizable with standard quadrotors.
When the desired thrust is negative, the controller drives the propeller speed to zero (a saturation state) in an attempt to achieve the thrust.
This action has two adverse effects.
Firstly and most obviously the tracking error increases significantly since the desired control effort is not available, and secondly the out-runner motors undergo an aggressive state change where they need to come to a complete halt and again instantaneously achieve a high RPM count.
This is not only strenuous for the motors and reduces their lifespan, it also is extremely expensive energy-wise reducing the available flight time of the UAV.
Similarly, if the desired thrust is too large and exceeds the motor capabilities, the controller forces the motor to operate in a suboptimal state at its maximum capacity, which is very strenuous, costly energy-wise, and again reduces the motor lifespan and the UAVs flight time.
Another important consideration is that the stability proofs accompanying the high level controllers do not account for thrust saturation; this is true for \cite{geomquadlee}, \cite{geomquadlee_asian}, \cite{qeopidfar} and for the majority of NCSs in the bibliography.
Thus to guarantee stability the desired control effort must be available, i.e., saturation must be avoided.

By studying the occurrence of negative thrusts through extensive simulations, it was observed that thrusts remain positive if the control task at hand is a position trajectory of a relatively reasonable rate.
On the other hand, if the control task entails a large angle attitude maneuver, the thrusts can certainly become negative, even if the attitude maneuver is conducted at very slow rates.
Therefore, it is important to develop a method, realizable in real time, that can distribute the computed control moment $^{b}\mathbf{u}$ to the four thrusters of the quadrotor, during attitude maneuvers, without interfering with the control objective while simultaneously complying with the following constraint,
\begin{IEEEeqnarray}{C}
f_{max}>f_{i}>f_{min},\>i=1,..,4\label{f_con}
\end{IEEEeqnarray}

This poses itself as a constrained optimization problem;
yet here we take advantage of the dynamics of the system, by utilizing a high level control strategy based on the concatenated use of two flight modes (a Position Mode and an Attitude Mode) as in \cite{geomquadlee}, \cite{geomquadlee_asian}, \cite{qeopidfar}, to allow us to develop an alternative solution that is extremely simple, fast and complies with the requirements stated above.

The solution starts with the realization that even though the quadrotor is underactuated in SE(3), it can be viewed as an overactuated platform in SO(3), the configuration space of its attitude dynamics.
Therefore for reasonable rate attitude maneuvers, this actuation redundancy allows one to achieve additional constraints.
A maneuver is characterized as one of \textit{reasonable rate} if it is realizable in the margin of the motor thrusts.
However exploiting this redundancy is not enough to avoid saturation during aggressive attitude tracking. A necessary building block to find a robust solution is to ensure that the control moment, $^{b}\mathbf{u}$, stays within the margins of the thrusters capabilities.
Next, we develop (a) a thrust allocation strategy, and (b) an attitude controller able to generate bounded, control moment $^{b}\mathbf{u}$, i.e. a solution that respects actuator constraints during aggressive attitude maneuvers.

\subsection{Thrust Allocation\label{thr_alocation}}
The control moment vector, $^{b}\mathbf{u}$, to be produced by the propellers is associated with the thrust vector, $\mathbf{F}\in\mathbb{R}^{4}$, by,
\begin{IEEEeqnarray}{rCl}
\mathbf{F}=\mathbf{A}^{\#}({}^b\mathbf{u})&,&\mathbf{A}=
\begin{bmatrix}
0&d&0&-d\\
-d&0&d&0\\
-b_{T}&b_{T}&-b_{T}&b_{T}\\
\end{bmatrix}\IEEEeqnarraynumspace\label{alloc}\\
\mathbf{A}^{\#}&=&\mathbf{A}^{T}(\mathbf{A}\mathbf{A}^{T})^{-1}\in\mathbb{R}^{4\times3}
\label{pinv}
\end{IEEEeqnarray}
where $\mathbf{A}$ always has full row rank, and $\mathbf{A}^{\#}$ is the Moore-Penrose pseudoinverse.
The null space of (\ref{alloc}) can be exploited to achieve additional tasks by allocating $^{b}\mathbf{u}$ using, 
\begin{IEEEeqnarray}{L}
\mathbf{F}{=}\mathbf{A}^{\#}({}^b\mathbf{u}){+}(\mathbf{I}{-}\mathbf{A}^{\#}\mathbf{A})\boldsymbol{\xi}\IEEEyesnumber
\label{nullspa}
\end{IEEEeqnarray}
where $\boldsymbol{\xi}{\in}\mathbb{R}^{4}$ is a suitable vector designed to avoid saturations as a first priority and secondly to allow the quadrotor during the attitude maneuver to stay ''close'' to a desired position.
Based on the design specifications, $\boldsymbol{\xi}$ is defined as,
\begin{IEEEeqnarray}{rCL}
\boldsymbol{\xi}&{=}&\int^{t2}_{t1}\!\!{\nabla}_{\mathbf{F}}H(\mathbf{F})d\tau{+}{\begin{bmatrix}
1&1&1&1\\
0&d&0&-d\\
-d&0&d&0\\
-b_{T}&b_{T}&-b_{T}&b_{T}\\
\end{bmatrix}}^{\!-1}
\!{\begin{bmatrix}
f_{p}\\
0\\
0\\
0\\
\end{bmatrix}}\IEEEyesnumber
\label{xi}\\
f_{p}&{=}&{\left(\!{\begin{bmatrix}
\iota_{x}&0&0\\
0&\iota_{y}&0\\
0&0&\iota_{z}\\
\end{bmatrix}}\!{(mg\mathbf{E}_{3}+k_{\xi}({-}k_{v}\mathbf{e}_{v}{-}k_{x}\mathbf{e}_{x}){+}m\ddot{\mathbf{x}}_{d})}\!\!\right)}^{\!\!\!T}\!\!\mathbf{R}\mathbf{e}_{3}\IEEEnonumber
\end{IEEEeqnarray}

The first term of (\ref{xi}) applies actuator constraints by the gradient of a suitable function, $H(\mathbf{F}){=}\sum_{i=1}^{4}h(f_{i})$, \cite{Ryll}, with,
\begin{IEEEeqnarray*}{L}
h(f_{i})=
\begin{cases}
k_{h_{1}}tan^{2}(\frac{\pi(\lvert f_{i}\rvert-f_{idl})}{2(f_{idl}-f_{min})}), & f_{min}{<}\lvert f_{i}\rvert{\leq}f_{idl}\\
\frac{k_{h_{2}}}{2}(\lvert f_{i}\rvert-f_{idl})^{2}+\frac{(\lvert f_{i}\rvert-f_{idl})^{2}}{(\lvert f_{i}\rvert-f_{max})},& \lvert f_{i}\rvert>f_{idl}
\end{cases}
\end{IEEEeqnarray*}
$f_{min},f_{idl},f_{max}\in\mathbb{R}^{+}$ are the minimum, idle, and maximum thrusts respectively and $k_{h_{1}},k_{h_{2}}\in\mathbb{R}^{+}$ are tuning gains.
The action of $H(\mathbf{F})$ keeps $f_{i}$ as close to $f_{idl}$ and between $f_{min}$ and $f_{max}$.
Through the definition of $h(f_{i})$ the actuator constraints objective, implicitly has a higher priority than the position tracking, because $h(f_{i}){\to}\infty$ if $f_{i}{\to}{}f_{min}$ or $f_{i}{\to}{}f_{max}$.
Thus the position tracking objective is realized strictly in the margins allowed by the actuator constraints.

The second term of (\ref{xi}) projects to the null-space a reference expression for the thrust magnitude $f_p$, which tracks a desired quadrotor position, $\mathbf{x}_{d}$.
The collective thrust from \cite{geomquadlee} is used, by pre-multiplying by a gain matrix its feedback components to assign different weights to each axis depending on the maneuver.
The gain, $k_{\xi}$, is needed to adjust/scale the influence of the position/velocity error vectors because as mentioned above, position tracking is performed strictly in the margins allowed by the actuator constraints.
Hence it is advised strongly that the initial position error is small at the beginning of the attitude maneuver.

Because both the position and actuator constraint objectives are projected through $\boldsymbol{\xi}$ to the null-space of $\mathbf{A}^{\#}$, it is ensured that the attitude control objective is unobstructed, assuring that the guarantees, i.e., notions of stability and regions of attraction, produced by the soon to be introduced stability proof, are valid during the attitude maneuver.
In this way, for reasonable rate maneuvers, (\ref{f_con}) holds always.

Note that the above solution is extremely fast to compute and implement in real time because $\mathbf{A}^{\#}$, ${\nabla}_{\mathbf{F}}H(\mathbf{F})$, and the inverse matrix in the second component of (\ref{xi}) can be computed in an analytic form off-line.
Consequently during implementation, the on-board microcontroler only needs to evaluate the precomputed analytic expressions.
We emphasize that the developed allocation strategy can only be used for tracking and not for large step changes in attitude, to avoid the generation of irregular outputs, a consequence of employing null-space projection methods.

\subsection{Attitude tracking control for aggressive maneuvers\label{conmodatt}}

\newcounter{Prop1}
\addtocounter{Prop1}{1}
\newcounter{Prop2}
\addtocounter{Prop2}{2}
\newcounter{Prop3}
\addtocounter{Prop3}{3}
\newcounter{Prop4}
\addtocounter{Prop4}{4}
\newcounter{Prop5}
\addtocounter{Prop5}{5}
\newcounter{sub}
\addtocounter{sub}{1}

Next, an attitude control system is developed, able to follow an arbitrary smooth desired orientation $\mathbf{R}_{d}(t)\in\text{SO(3)}$ and its associated angular velocity $^{b}\boldsymbol{\omega}_{d}(t)\in\mathbb{R}^{3}$ by generating a bounded control torque, $^{b}\mathbf{u}$.

\subsubsection{Attitude tracking errors}
For a given tracking command ($\mathbf{R}_{d}$, $^{b}\boldsymbol{\omega}_{d}$) and the current attitude and angular velocity ($\mathbf{R}$, $^{b}\boldsymbol{\omega}$), an \textit{attitude error function} $\Psi:\text{SO(3)}\times\text{SO(3)}\rightarrow\mathbb{R}$, an \textit{attitude error vector} $\mathbf{e}_{R}\in\mathbb{R}^{3}$, and an \textit{angular velocity error vector}, $\mathbf{e}_{\omega}\in\mathbb{R}^{3}$, are defined as follows, \cite{err_fun}:
\begin{IEEEeqnarray}{rCl}
\Psi(\mathbf{R},\mathbf{R}_{d})&=&2-\sqrt{1+\text{tr}[\mathbf{R}^{T}_{d}\mathbf{R}]}\geq 0
\label{error_function_A}
\end{IEEEeqnarray}
\begin{IEEEeqnarray}{rCl}
\mathbf{e}_{R}(\mathbf{R},\mathbf{R}_{d})&=&\frac{S^{-1}(\mathbf{R}^{T}_{d}\mathbf{R}-\mathbf{R}^{T}\mathbf{R}_{d})}{2\sqrt{1+tr[\mathbf{R}^{T}_{d}\mathbf{R}]}}\label{att_error_A}
\end{IEEEeqnarray}
\begin{IEEEeqnarray}{rCl}
\mathbf{e}_{\omega}(\mathbf{R},{}^{b}\boldsymbol{\omega},\mathbf{R}_{d},{}^{b}\boldsymbol{\omega}_{d})&=&{}^{b}\boldsymbol{\omega}-\mathbf{R}^{T}\mathbf{R}_{d}{}^{b}\boldsymbol{\omega}_{d}
\label{ang_vel_error}
\end{IEEEeqnarray}
where $\text{tr}[.]$ is the trace function.
Note that the maximum attitude difference, that of 180$^{o}$ with respect to an equivalent axis-angle rotation between $\mathbf{R}$ and $\mathbf{R}_{d}$, occurs when the rotation matrices are antipodal and at that instant (\ref{error_function_A}) yields $\Psi(\mathbf{R},\mathbf{R}_{d})$=2, i.e. 100\% error.
If both rotation matrices express the same attitude i.e., $\mathbf{R}$=$\mathbf{R}_{d}$, then $\Psi(\mathbf{R},\mathbf{R}_{d})$=0, i.e. 0\% error.
Important properties regarding (\ref{error_function_A})-(\ref{ang_vel_error}), including the associated attitude error dynamics used in this work are included in Propositions \arabic{Prop1} and \arabic{Prop2}  found in Appendix \ref{appA}.

\subsubsection{Attitude tracking controller}
Next a control system is defined using elements from the controllers in \cite{qeopidfar}, \cite{err_fun}, and additional modifications to ensure that the generated control effort remains bounded.
The developed control system stabilizes the attitude dynamics of the quadrotor UAV $\mathbf{e}_{R}$, $\mathbf{e}_{\omega}$, to zero exponentially, almost globally.

\textbf{Proposition \arabic{Prop3}.} 
For $k_{R},k_{\omega}{\in}\mathbb{R}^{+}$, and initial conditions,
\begin{IEEEeqnarray}{C}
\Psi(\mathbf{R}(0),\mathbf{R}_{d}(0))<2
\label{Psi_0}\\
\lVert\mathbf{e}_{\omega}(0)\rVert^{2}{<}\frac{2 k_{R}}{\lambda_{max}(\mathbf{J})}\left(2{-}\Psi(\mathbf{R}(0),\mathbf{R}_{d}(0))\right)\label{surface_0}
\end{IEEEeqnarray}
and for sufficiently smooth desired attitude $\mathbf{R}_{d}(t){\in}\text{SO(3)}$ in,
\begin{IEEEeqnarray}{rCl}
L_{2}&=&\{(\mathbf{R},\mathbf{R}_{d})\in\text{SO(3)}\times\text{SO(3)}|\Psi(\mathbf{R},\mathbf{R}_{d})<2\}
\label{L_2}
\end{IEEEeqnarray}
such that for a chosen $B_{2}\in\mathbb{R}^{+}$ the following is valid,
\begin{IEEEeqnarray}{rCl}
\lVert 2\mathbf{J}-\text{tr}[\mathbf{J}]\mathbf{I}\rVert \lVert{}^{b}\boldsymbol{\omega}_{d}\rVert&\leq&B_{2} \label{cond_B_2}
\end{IEEEeqnarray}
we define the following controller,
\begin{IEEEeqnarray}{rCl}
^{b}\mathbf{u}&=&-k_{R}\mathbf{e}_{R}-k_{\omega}\frac{\mathbf{e}_{\omega}}{\sqrt{1+\mathbf{e}_{\omega}^{T}\mathbf{e}_{\omega}}}+\mathbf{J}\mathbf{R}^{T}\mathbf{R}_{d}{}^{b}\dot{\boldsymbol{\omega}}_{d}\label{att_contr}\\
&&+S(\mathbf{R}^{T}\mathbf{R}_{d}{}^{b}\boldsymbol{\omega}_{d})\mathbf{J}\mathbf{R}^{T}\mathbf{R}_{d}{}^{b}\boldsymbol{\omega}_{d}\IEEEnonumber
\end{IEEEeqnarray}
Then the zero equilibrium of the quadrotor closed loop attitude tracking error $(\mathbf{e}_{R},\mathbf{e}_{\omega})=(\mathbf{0},\mathbf{0})$ is almost globally exponentially stable;
moreover there exist constants $\mu,\tau>0$ such that
\begin{IEEEeqnarray}{C}
\Psi(\mathbf{R},\mathbf{R}_{d})<min\{2,\mu e^{-\tau t}\}
\label{Psi_bou}
\end{IEEEeqnarray}

\textbf{Proof}:
First we show that given the conditions (\ref{Psi_0})-(\ref{surface_0}), $L_{2}$ is positively invariant.
Then we show almost global exponential stability of the attitude error dynamics in $L_{2}$.
\begin{enumerate}[(a)]
\item Boundedness of $\Psi$: We use the Lyapunov function, \cite{err_fun},
\begin{IEEEeqnarray}{rCl}
V_{\Psi}&=&\frac{1}{2}\mathbf{e}_{\omega}^{T}\mathbf{J}\mathbf{e}_{\omega}+k_{R}\Psi\label{ater_lyap}
\end{IEEEeqnarray}
Differentiating (\ref{ater_lyap}), substituting App. \ref{appA}(\ref{att_error_dyn}), (\ref{att_contr}) we get,\begin{IEEEeqnarray}{rCl}
\dot{V}_{\Psi}&=&-k_{\omega}\frac{\lVert\mathbf{e}_{\omega}\rVert^{2}}{\sqrt{1+\mathbf{e}_{\omega}^{T}\mathbf{e}_{\omega}}}\leq0
\label{dater_lyap}
\end{IEEEeqnarray}
 Equations (\ref{ater_lyap}-\ref{dater_lyap}) imply that for $\lVert\mathbf{e}_{\omega}\rVert<\infty$, $V_{\Psi}(t)\leq V_{\Psi}(0),\forall t\geq 0$.
Applying (\ref{surface_0}) we obtain,
\begin{IEEEeqnarray}{C}
k_{R} \Psi(\mathbf{R}(t),\mathbf{R}_{d}(t)){\leq} V_{\Psi}(t){\leq} V_{\Psi}(0){<}2 k_{R} \label{hkrko_Psi}
\end{IEEEeqnarray}
implying that the attitude error function is bounded by,
\begin{IEEEeqnarray}{C}
\Psi(\mathbf{R}(t),\mathbf{R}_{d}(t))\leq \psi_{a} < 2,\;\forall \;t\geq 0\label{B_Psi}
\end{IEEEeqnarray}
where $\psi_{a}={V_{\Psi}(0)}/{k_{R} }$. 
Thus under the aforementioned conditions $L_{2}$ is positively invariant with $\mathbf{R}(t){\in} L_{2},\forall\;t$.
\item Boundedness of $\mathbf{e}_{\omega}$: Using (\ref{ater_lyap}), (\ref{hkrko_Psi}) the following holds,
\begin{IEEEeqnarray}{L}
\frac{1}{2}\lambda_{min}(\mathbf{J})\lVert\mathbf{e}_{\omega}\rVert^{2}{\leq}\frac{1}{2}\mathbf{e}_{\omega}^{T}\mathbf{J}\mathbf{e}_{\omega}{\leq} V_{\Psi}(t){\leq} V_{\Psi}(0){<}2 k_{R} \label{hkrko_eo}\;\;\;
\end{IEEEeqnarray}
implying that the angular velocity error is bounded by,
\begin{IEEEeqnarray}{C}
\lVert\mathbf{e}_{\omega}(t)\rVert^{2}<\frac{4k_{R}}{\lambda_{min}(\mathbf{J})},\forall t\label{maxeo}
\end{IEEEeqnarray}
\item Lyapunov candidate: We use the Lyapunov function, \cite{qeopidfar},
\begin{IEEEeqnarray}{rCl}
V&=&V_{\Psi}+c_{2}\mathbf{e}_{R}^{T}\mathbf{J}\mathbf{e}_{\omega}
\label{att_lyap}
\end{IEEEeqnarray}
Differentiating (\ref{att_lyap}), inserting App. \ref{appA}(\ref{att_error_dyn}), (\ref{att_contr}), we get,
\begin{IEEEeqnarray}{rCl}
\dot{V}&=&-k_{\omega}\frac{\lVert\mathbf{e}_{\omega}\rVert^{2}}{\sqrt{1+\mathbf{e}_{\omega}^{T}\mathbf{e}_{\omega}}}+c_{2}\dot{\mathbf{e}}_{R}\cdot\mathbf{J}\mathbf{e}_{\omega}-c_{2}k_{R}\lVert\mathbf{e}_{R}\rVert^{2}\IEEEnonumber\\
&&-c_{2}k_{\omega}\frac{\mathbf{e}_{R}^{T}\mathbf{e}_{\omega}}{\sqrt{1+\mathbf{e}_{\omega}^{T}\mathbf{e}_{\omega}}}+c_{2}\mathbf{e}_{R}\cdot S(\mathbf{J}\mathbf{e}_{\omega}+\mathbf{d})\mathbf{e}_{\omega}
\label{Datt_lyap}
\end{IEEEeqnarray}
where $\mathbf{d}=(2\mathbf{J}-\text{tr}[\mathbf{J}]\mathbf{I})\mathbf{R}^{T}\mathbf{R}_{d}{}^{b}\boldsymbol{\omega}_{d}\in\mathbb{R}^{3}$.
Using (\ref{maxeo}) we define $B_{1}{=}1/\sqrt{1+(4 k_{R}/\lambda_{min}(\mathbf{J}))}$, thus, 
\begin{IEEEeqnarray}{rCl}
\dot{V}&\leq&-k_{\omega}B_{1}\lVert\mathbf{e}_{\omega}\rVert^{2}+c_{2}\dot{\mathbf{e}}_{R}\cdot\mathbf{J}\mathbf{e}_{\omega}-c_{2}k_{R}\lVert\mathbf{e}_{R}\rVert^{2}\IEEEnonumber\\
&&+c_{2}k_{\omega}\lVert\mathbf{e}_{R}\rVert\cdot\lVert\mathbf{e}_{\omega}\rVert{+}c_{2}\mathbf{e}_{R}\cdot S(\mathbf{J}\mathbf{e}_{\omega}+\mathbf{d})\mathbf{e}_{\omega}
\label{Datt_lyap_1}
\end{IEEEeqnarray}
Note that by (\ref{cond_B_2}), $\lVert\mathbf{d}\rVert{\leq} B_{2}$.
Using $\lVert\mathbf{e}_{R}\rVert{\leq}1$, App. \ref{appA}(\ref{erDot_norm_A}), 
\begin{IEEEeqnarray}{rCl}
\dot{V}&\leq&-\mathbf{z}^{T}_{R}\mathbf{W}_{2}\mathbf{z}_{R}
\label{Datt_lyap}
\end{IEEEeqnarray}
where $\mathbf{z}_{R}=[\lVert\mathbf{e}_{R}\rVert;\lVert\mathbf{e}_{\omega}\rVert]$ and $\mathbf{W}_{2}\in\mathbb{R}^{2\times2}$ is given by,
\begin{IEEEeqnarray}{rCl}
\mathbf{W}_{2}=\begin{bmatrix}
k_{R}c_{2}&-\frac{c_{2}}{2}(k_{\omega}+B_{2})\\
-\frac{c_{2}}{2}(k_{\omega}+B_{2})&k_{\omega}B_{1}-\frac{3}{2}c_{2}\lambda_{max}(\mathbf{J})
\end{bmatrix}\label{eq:W_2}
\end{IEEEeqnarray}

\item Exponential Stability: Using App. \ref{appA}(\ref{quadr_Psi}), $V$ is bounded,
\begin{IEEEeqnarray}{C}
\mathbf{z}^{T}_{R}\mathbf{\Pi}_{1}\mathbf{z}_{R}\leq V\leq\mathbf{z}^{T}_{R}\mathbf{\Pi}_{2}\mathbf{z}_{R}
\end{IEEEeqnarray}
where $\mathbf{\Pi}_{1}$, $\mathbf{\Pi}_{2}$ are given by, \cite{err_fun},
\begin{IEEEeqnarray}{rCl}
\mathbf{\Pi}_{1}=
\begin{bmatrix}
k_{R}&\frac{c_{2}}{2}\\
\frac{c_{2}}{2}&\frac{\lambda_{min}(\mathbf{J})}{2}
\end{bmatrix},
\mathbf{\Pi}_{2}=
\begin{bmatrix}
2k_{R}&\frac{c_{2}}{2}\\
\frac{c_{2}}{2}&\frac{\lambda_{max}(\mathbf{J})}{2}
\end{bmatrix}
\end{IEEEeqnarray}
then by choosing $c_{2}\in\mathbb{R}^{+}$ such that,
\begin{IEEEeqnarray}{rCl}
c_{2}<&\min\big\{&\sqrt{2k_{R}\lambda_{min}(\mathbf{J})},\frac{2k_{\omega}B_{1}}{3\lambda_{max}(\mathbf{J})},\IEEEnonumber\\
&&\frac{4k_{R}k_{\omega}B_{1}}{6k_{R}\lambda_{max}(\mathbf{J})+(k_{\omega}+B_{2})^{2}}\big\}\IEEEyesnumber\label{eq:c_1}
\end{IEEEeqnarray}
the matrices $\mathbf{\Pi}_{1}$, $\mathbf{\Pi}_{2}$, $\mathbf{W}_{2}$, are positive definite.
Thus the following inequalities are valid,
\begin{IEEEeqnarray}{C}
\lambda_{min}(\mathbf{\Pi}_{1})\lVert \mathbf{z}_{R} \rVert^{2}\leq V \leq\lambda_{max}(\mathbf{\Pi}_{2})\lVert\mathbf{z}_{R}\rVert^{2}\\
\dot{V} \leq -\lambda_{min}(\mathbf{W}_{2})\lVert\mathbf{z}_{R}\rVert^{2}
\end{IEEEeqnarray}
Then for $\tau=\frac{\lambda_{min}(\mathbf{W}_{2})}{\lambda_{max}(\mathbf{\Pi}_{2})}$ the following holds,
\begin{IEEEeqnarray}{C}
\dot{V} \leq -\tau V
\end{IEEEeqnarray}
Thus the zero equilibrium of the attitude tracking error $\mathbf{e}_{R}$, $\mathbf{e}_{\omega}$ is exponentially stable almost globally.
Finally, using App. \ref{appA}(\ref{quadr_Psi}) then,
\begin{IEEEeqnarray}{C}
\frac{1}{2}\lambda_{min}(\mathbf{\Pi}_{1})\Psi \leq  V(t) \leq V(0)e^{-\tau t}\IEEEyesnumber
\end{IEEEeqnarray}
Thus $\Psi$ exponentially decreases and therefore (\ref{B_Psi}) implies (\ref{Psi_bou}).
This completes the proof. $\blacksquare$
\end{enumerate}

The region of attraction given by (\ref{Psi_0})-(\ref{surface_0}) ensures that the initial attitude error is less than $180^{o}$ wrt. an axis-angle rotation (i.e., $\mathbf{R}_{d}(t)$ is not antipodal to $\mathbf{R}(t)$).
Consequently, exponential stability is guaranteed almost globally (everywhere except the antipodal equilibrium).
This is the best one can do since it was shown that the topology of SO(3) prohibits the design of a smooth global controller \cite{obstruction}.

The selection of a suitable $B_{2}$ for (\ref{cond_B_2}), ensures that the rate of change of the reference attitude trajectory is gradual/slow enough to be negotiated by the motors without the emergence of saturations (this is achieved by trajectory design such that (\ref{cond_B_2}) holds).
Finally, provided that the reference attitude trajectory ($\mathbf{R}_{d}(t)$, ${}^{b}\boldsymbol{\omega}_{d}(t)$, ${}^{b}\dot{\boldsymbol{\omega}}_{d}(t)$), is sufficiently smooth and bounded, all the terms in (\ref{att_contr}) are bounded.
Thus, through proper gain selection, (\ref{att_contr}) can be tuned to remain in the allowable margins dictated by the actuator constraints.

Concluding, since (\ref{att_contr}) is developed directly on SO(3), it completely avoids singularities and ambiguities associated with minimum attitude representations like Euler angles or quaternions.
Also the controller can be applied to the attitude dynamics of any rigid body and not only on quadrotor systems.

\section{Stability of the Position Control Mode\label{conmodpos}}

Next we prove stability for a position controler composed by the developed attitude controller, (\ref{att_contr}), and the collective thrust expression from \cite{geomquadlee}.

\subsection{Position tracking controller\label{sec:thr_mag}}
For a sufficiently smooth pointing trajectory $\mathbf{e}_{1_{d}}(t)\in\text{S}^{2}=\{\mathbf{q}\in\mathbb{R}^{3}|\mathbf{q}^{T}\mathbf{q}=1\}$ associated with the yaw orientation of the quadrotor UAV, and tracking instruction $\mathbf{x}_{d}(t)\in\mathbb{R}^{3}$, a position controller is defined, composed by the developed attitude controller, (\ref{att_contr}), and the collective thrust expression from \cite{geomquadlee}, given by,
\begin{IEEEeqnarray}{rCL}
f&=&(mg\mathbf{E}_{3}{-}k_{v}\mathbf{e}_{v}{-}k_{x}\mathbf{e}_{x}{+}m\ddot{\mathbf{x}}_{d})^{T}\mathbf{R}\mathbf{e}_{3}\label{f}
\end{IEEEeqnarray}
where $\mathbf{e}_{x}$, $\mathbf{e}_{v}$, are the position and velocity tracking errors,
\begin{IEEEeqnarray}{C}
\mathbf{e}_{x}=\mathbf{x}-\mathbf{x}_{d},\; \mathbf{e}_{v}=\mathbf{v}-\dot{\mathbf{x}}_{d}\label{pos_error}
\end{IEEEeqnarray}
The closed loop system is  defined by (\ref{eq:position})-(\ref{attitude_kinem}) under the action of (\ref{att_contr}), (\ref{f}), and is shown to achieve almost global exponential stabilization of ($\mathbf{e}_{x}$,$\mathbf{e}_{v}$,$\mathbf{e}_{R}$,$\mathbf{e}_{\omega}$) to zero next. 

\textbf{Proposition \arabic{Prop4}.} 
Consider the developed attitude controller, (\ref{att_contr}), and the collective thrust expression from \cite{geomquadlee}, given by (\ref{f}), with initial conditions in (\ref{Psi_0}), (\ref{surface_0}), and suitable gains $k_{R}$, $k_{\omega}$.
Then for a smooth yaw pointing trajectory $\mathbf{e}_{1_{d}}(t)\in\text{S}^{2}$, and position tracking instruction $\mathbf{x}_{d}(t)\in\mathbb{R}^{3}$ such that (\ref{cond_B_2}) holds, the zero equilibrium for the complete system, i.e., $\mathbf{e}_{x}$, $\mathbf{e}_{v}$, $\mathbf{e}_{R}$, $\mathbf{e}_{\omega}$, is exponentially stable, almost globally.

\textbf{Proof}:
See Proposition 2 in \cite{geomquadlee}, but replace Eqs. (20), (23), (25), (27) and $\alpha{\in}\mathbb{R}^{+}$ in \cite{geomquadlee}, with Eqs. (\ref{B_Psi}), (\ref{eq:W_2}), (\ref{eq:c_1}), (\ref{surface_0}), $\alpha{=}\sqrt{\psi_{a}(1{-}{\psi_{a}/4})}$ respectively, from Section \ref{conmodatt} here.

Note that the gains $k_{R}$, $k_{\omega}$, must be chosen according to Proposition \arabic{Prop4}.
Also during position tracking, the attitude dynamics are driven to track a computed attitude, $\mathbf{R}_{x}(t)\in\text{SO}(3)$, constructed based on the desired yaw pointing trajectory $\mathbf{e}_{1_{d}}(t)\in\text{S}^{2}$, and tracking instruction $\mathbf{x}_{d}(t)\in\mathbb{R}^{3}$.
The procedure on performing this computation is given in \cite{geomquadlee_asian}.



\subsection{Quadrotor Tracking Controls\label{tracControls}}
Concluding, in this work two flight modes are utilized:
\begin{itemize}
\item \textit{Attitude Mode}:
The controller achieves tracking for the attitude of the quadrotor UAV while avoids saturations and stays ''close'' to a desired quadrotor position, through the combined action of (\ref{nullspa}), (\ref{xi}), (\ref{att_contr}).
\item \textit{Position Mode}:
The controller achieves tracking of a smooth position instruction, $\mathbf{x}_{d}(t){\in}\mathbb{R}^{3}$, for the quadrotor CM, and a pointing trajectory, $\mathbf{e}_{1_{d}}(t){\in}\text{S}^{2}$, associated with the yaw orientation of the quadrotor UAV through the combined action of (\ref{att_contr}) and (\ref{f}).
\end{itemize}

Using these flight modes in suitable successions, a quadrotor can perform a complex desired flight maneuver.
The fact that the region of attraction of each mode is almost global (see Prop. \arabic{Prop3} and Prop. \arabic{Prop4}) allows the safe switching between flight modes. 
It is emphasized that the attitude mode is better suited for short durations of time because during the Attitude Mode, the quadrotor stays ''close'' to a desired quadrotor position as a secondary objective wrt. the attitude tracking instruction and does so only in the margins dictated by the actuator constraints.

\section{Simulations\label{simulation}}

To assess the developed controllers' effectiveness an aggressive maneuver will be negotiated using the composed flight modes.
The simulation will be conducted twice:
once with the controllers using the developed allocation strategy during the Attitude Mode, nicknamed Null-space solution, and once with the controllers using the solution from \cite{geomquadlee}, i.e., a collective thrust expression to track a desired altitude command, nicknamed Benchmark solution.
The system parameters were taken from a real quadrotor described in \cite{Hinf}:
\begin{IEEEeqnarray*}{C}
\mathbf{J}=[0.0181,0,0;0,0.0196,0;0,0,0.0273]\;kg\,m^{2}\IEEEnonumber\\
m=1.225\;kg,
d=0.23\;m,
b_{T}=0.0121\;m\IEEEnonumber
\end{IEEEeqnarray*}
and the actuator constraints, see \cite{Hinf}, are given by:
\begin{IEEEeqnarray*}{C}
f_{i,min}=0{ }\text{[N]}, f_{i,max}=6.9939{ }\text{[N]}
\end{IEEEeqnarray*}
All the simulations were conducted using fixed-step integration with $dt{=}1{\cdot}10^{-3}$s.
The controller parameters are: 
\begin{IEEEeqnarray*}{C}
k_{\omega}=[2.172,0,0;0,2.352,0;0,0,3.276]\\
k_{R}{=}[65.16,0,0;0,70.56,0;0,0,98.28]\\
k_{v}{=}48.6521,k_{x}{=}453.6205\\
k_{h_{1}}{=}2,k_{h_{2}}{=}3,\iota_{x}{=}1.5,\iota_{y}{=}1,\iota_{z}{=}1.25,k_{\xi}{=}0.05
\end{IEEEeqnarray*}

A complex flight maneuver is conducted, involving several transitions between flight modes.
The initial conditions (IC's) are: $\mathbf{x}(0){=}\mathbf{v}(0){=}{}^{b}\boldsymbol{\omega}(0){=}\mathbf{0}_{3\times 1},\mathbf{R}(0){=}\mathbf{I}$.
The trajectory to be achieved through the concatenation of the two flight modes is:
\begin{enumerate}[(a)]
\item ($t < 6$): Position Mode: At $t=0.5$s the quadrotor translates from the origin to $\mathbf{x}_{d}=[2;0;10],\mathbf{e}_{1d}=[1;0;0]$ using Smooth Polynomials of eighth degree (SP$8^{th}$).
\item ($6\leq t < 7$): Attitude Mode: The quadrotor performs a $360^{o}$ flip around its $\mathbf{e}_{2}$ axis.
$\mathbf{R}_{d}(t)$ was designed by defining the pitch angle using SP$8^{th}$.
\item ($7\leq t \leq 10$): Trajectory tracking using SP$8^{th}$ with IC's equal to the values of the states of the quadrotor at the end of the flip and final waypoint given by $\mathbf{x}_{d}=[2;0;10],\mathbf{e}_{1d}=[1;0;0]$.
\end{enumerate}

\begin{figure}[!h]
\centering
\subfloat[\label{PsiComp}]{\includegraphics[width=0.5\columnwidth]{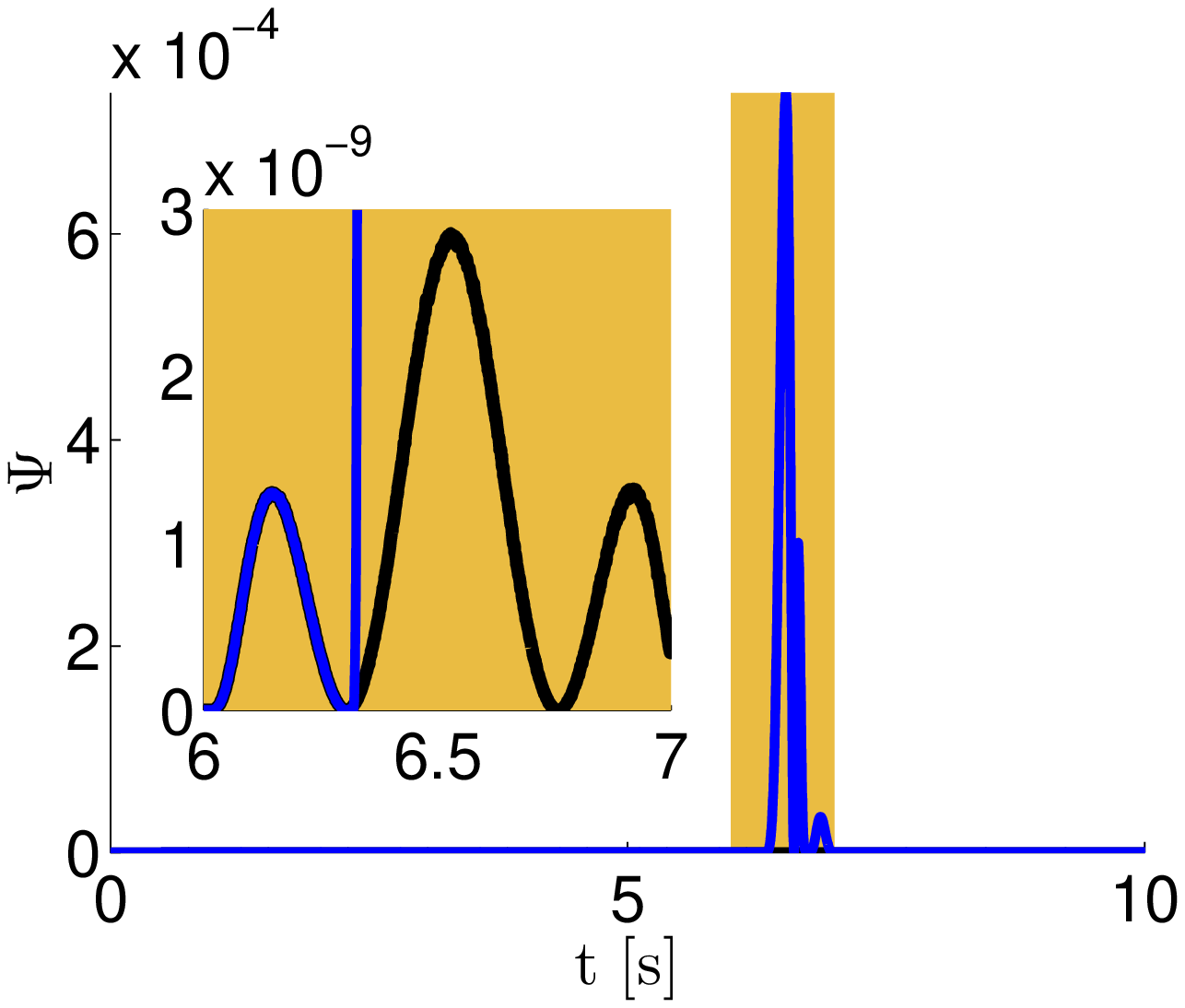}}
~
\subfloat[\label{normgon}]{\includegraphics[width=0.5\columnwidth]{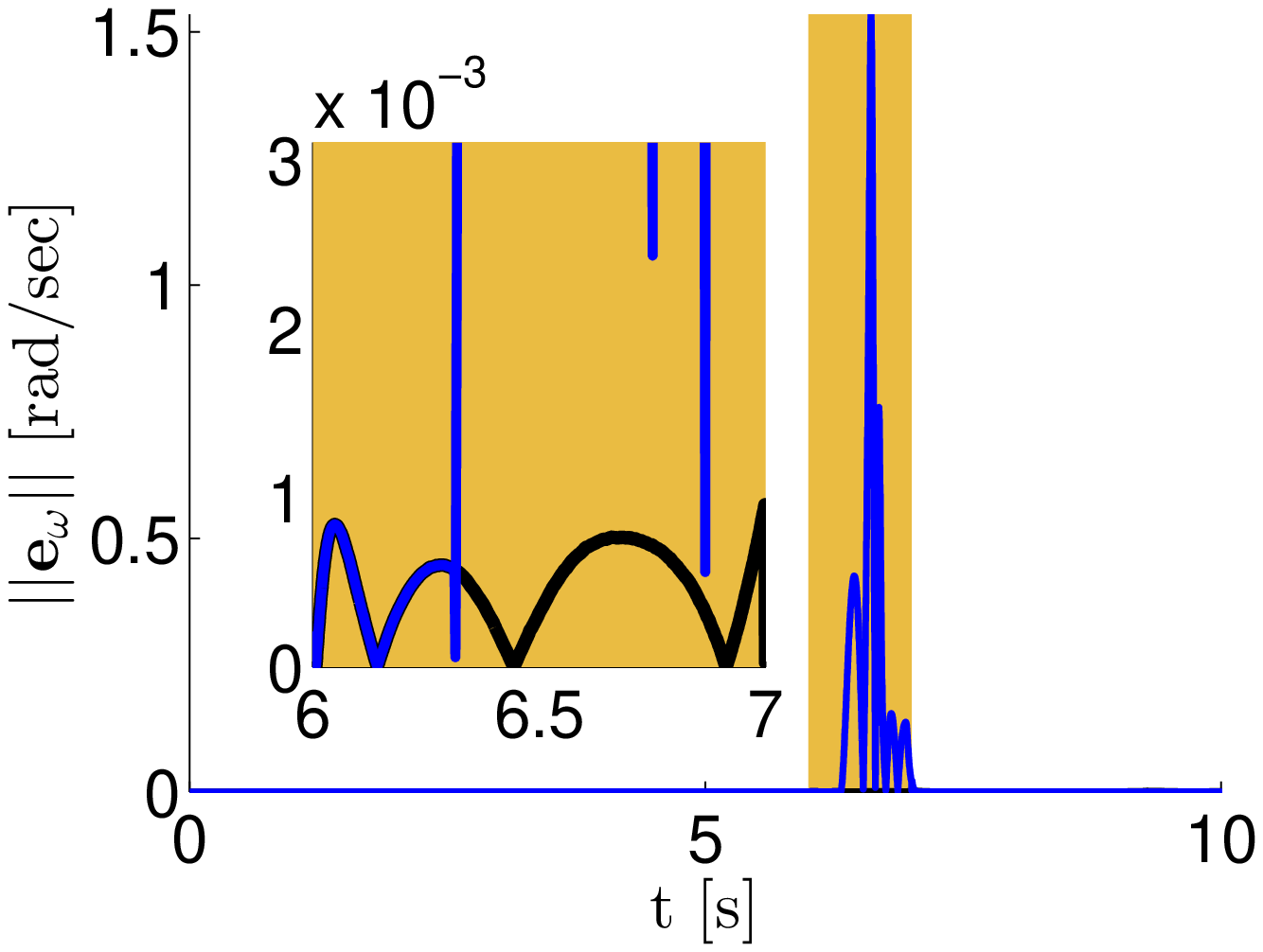}}

\subfloat[\label{normpos}]{\includegraphics[width=0.5\columnwidth]{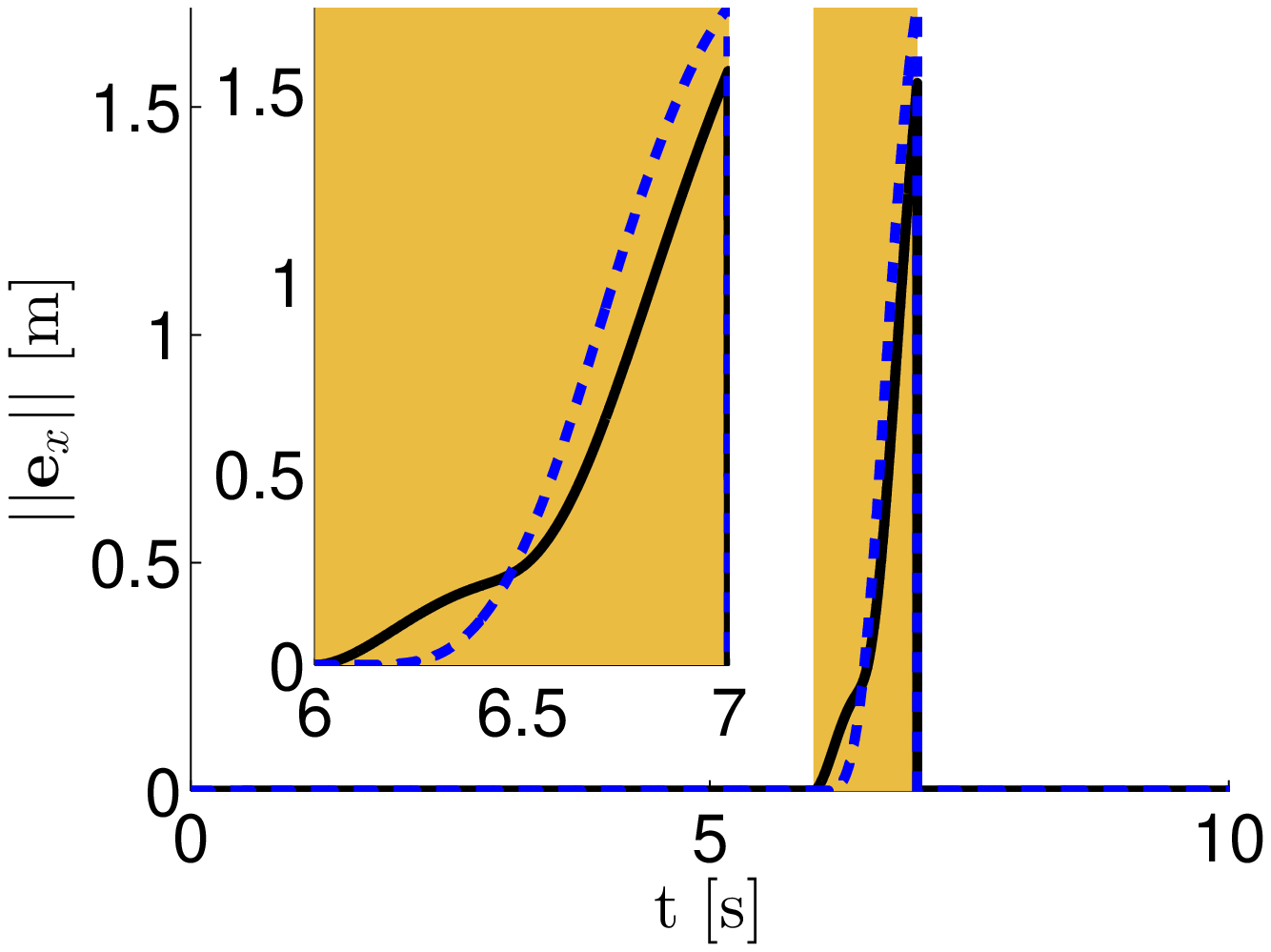}}
~
\subfloat[\label{trajU}]{\includegraphics[width=0.5\columnwidth]{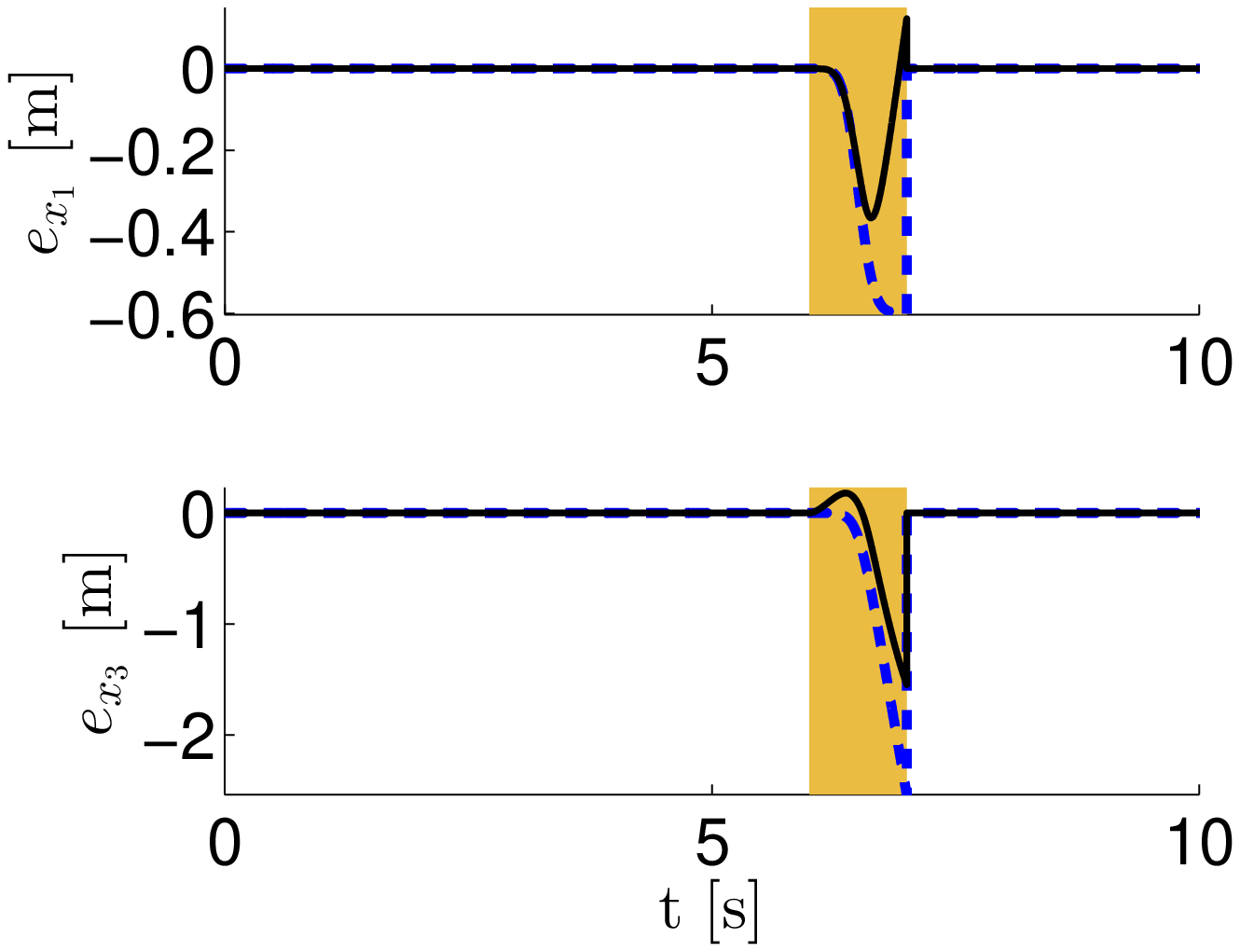}}

\subfloat[\label{thrR11}]{\includegraphics[width=0.5\columnwidth]{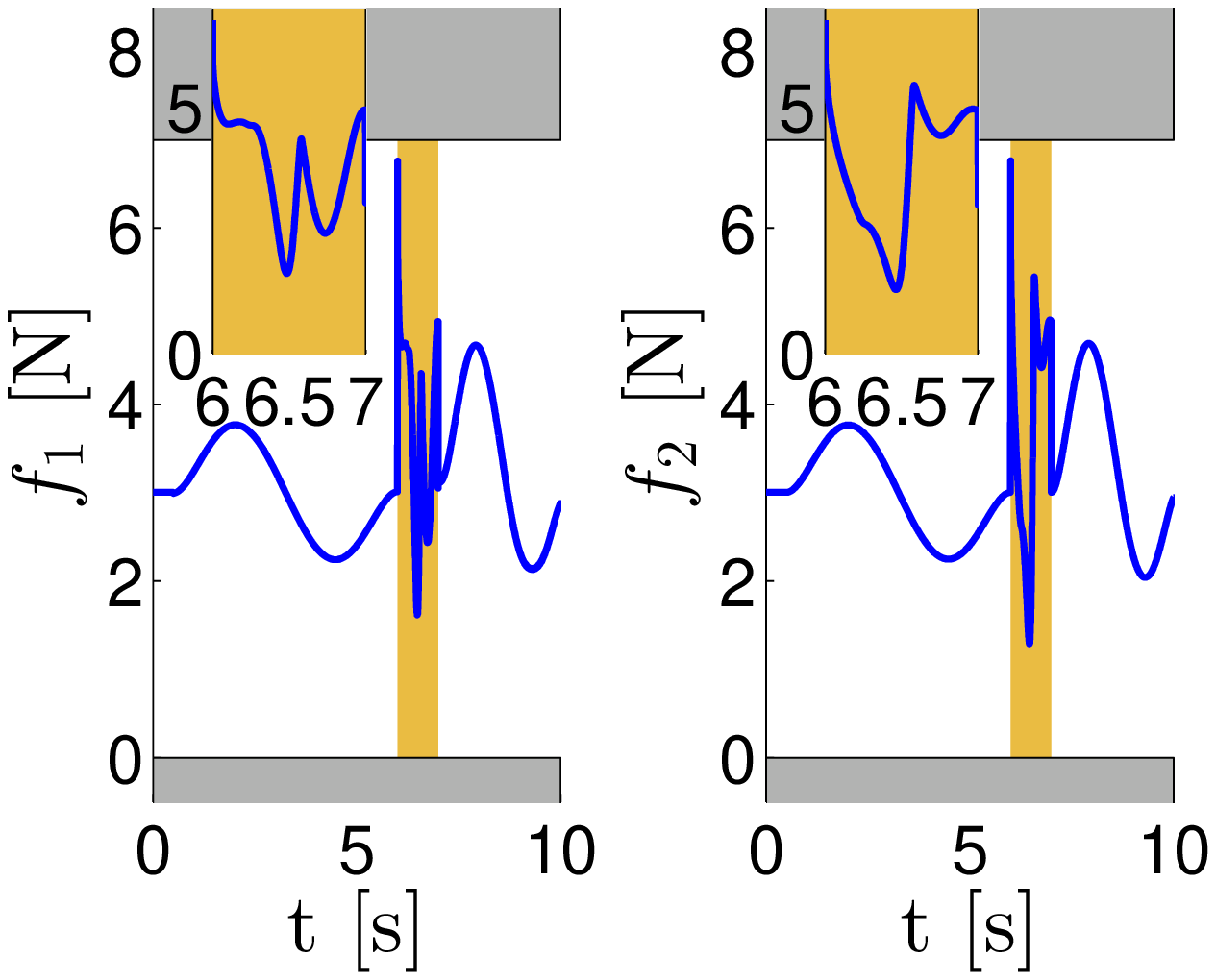}}
~
\subfloat[\label{thrR12}]{\includegraphics[width=0.5\columnwidth]{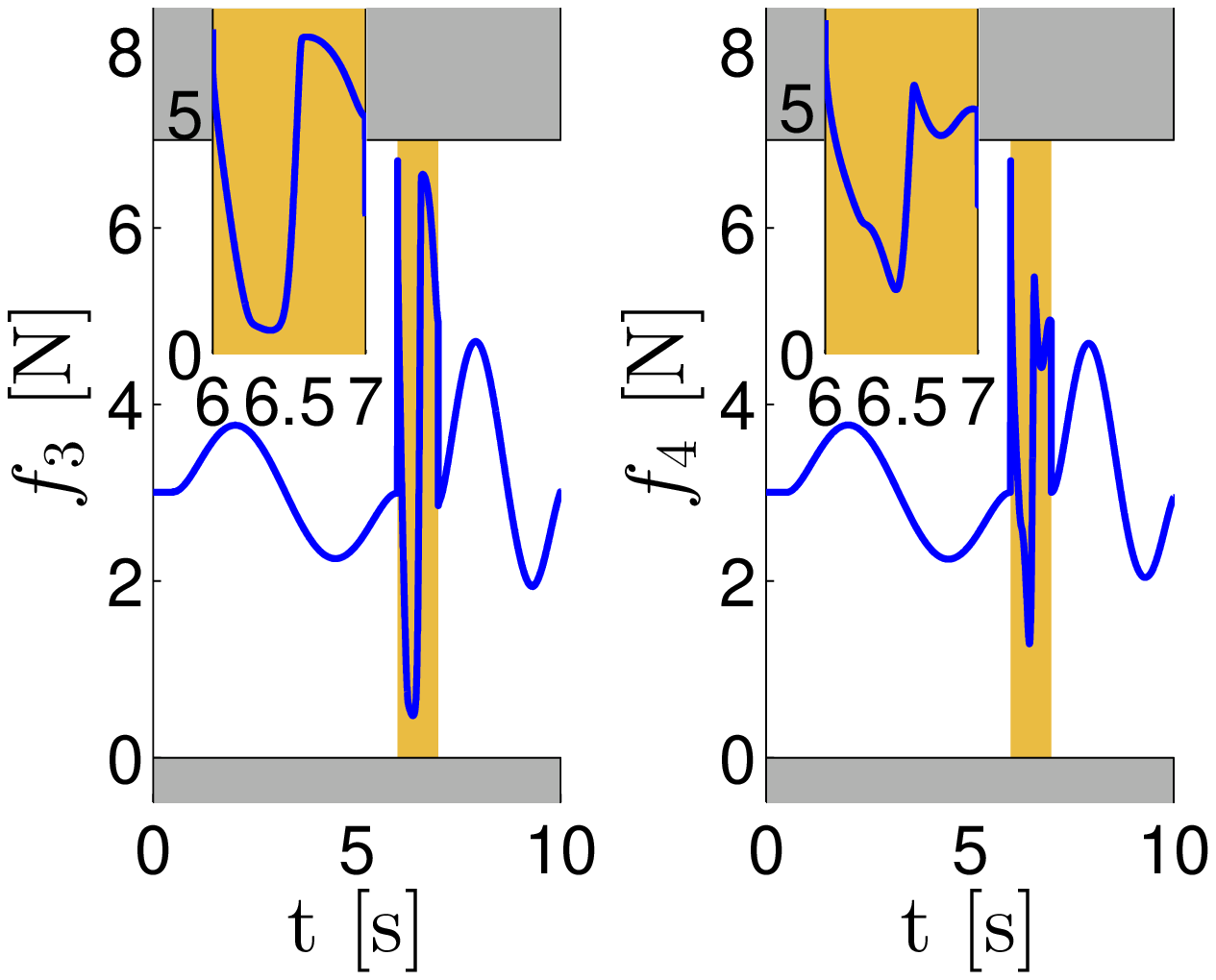}}

\subfloat[\label{thrM11}]{\includegraphics[width=0.5\columnwidth]{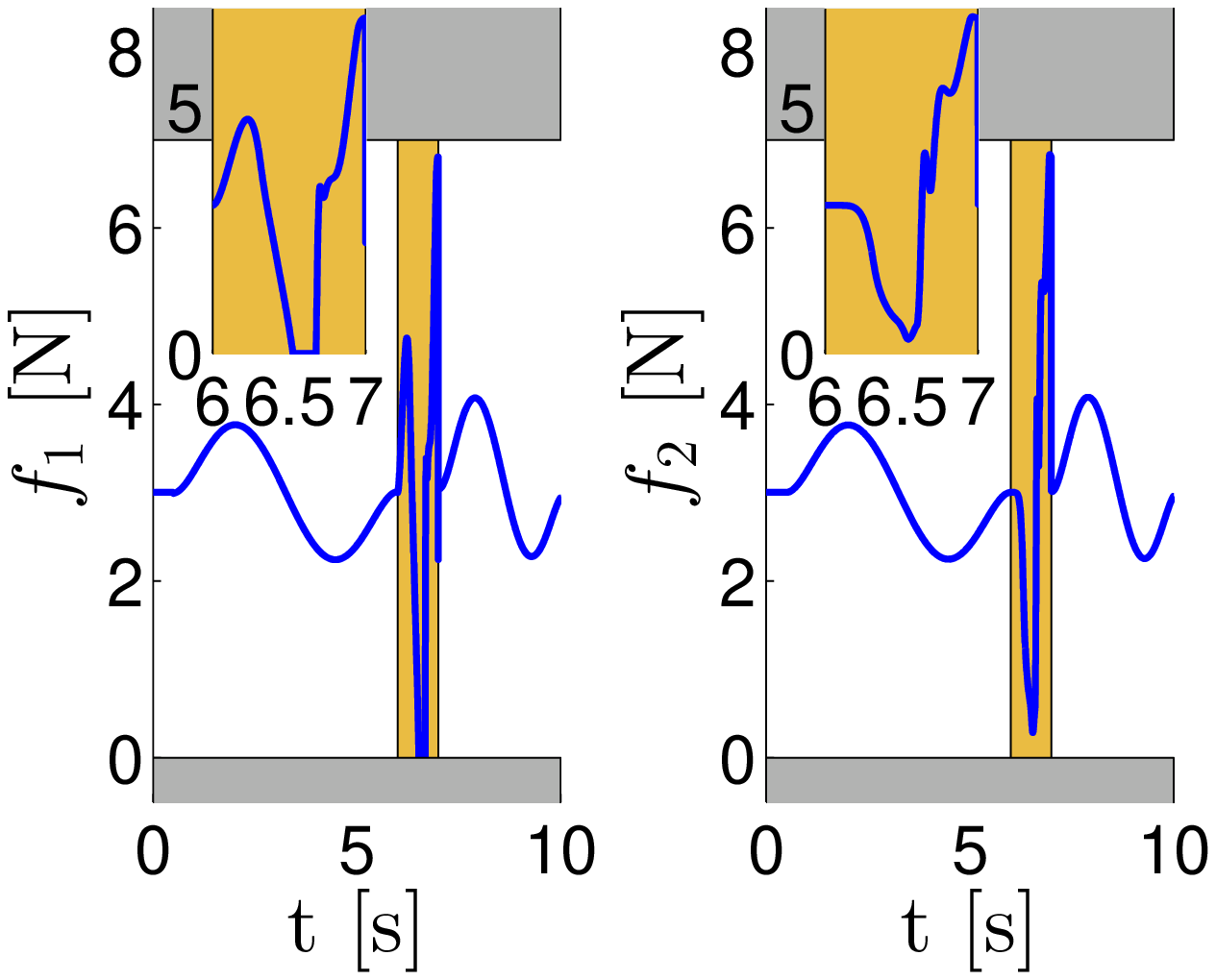}}
\put(-94,25){\parbox{\columnwidth}{${\searrow}$}}
\put(-104,32){\parbox{\columnwidth}{sat}}
~
\subfloat[\label{thrM12}]{\includegraphics[width=0.5\columnwidth]{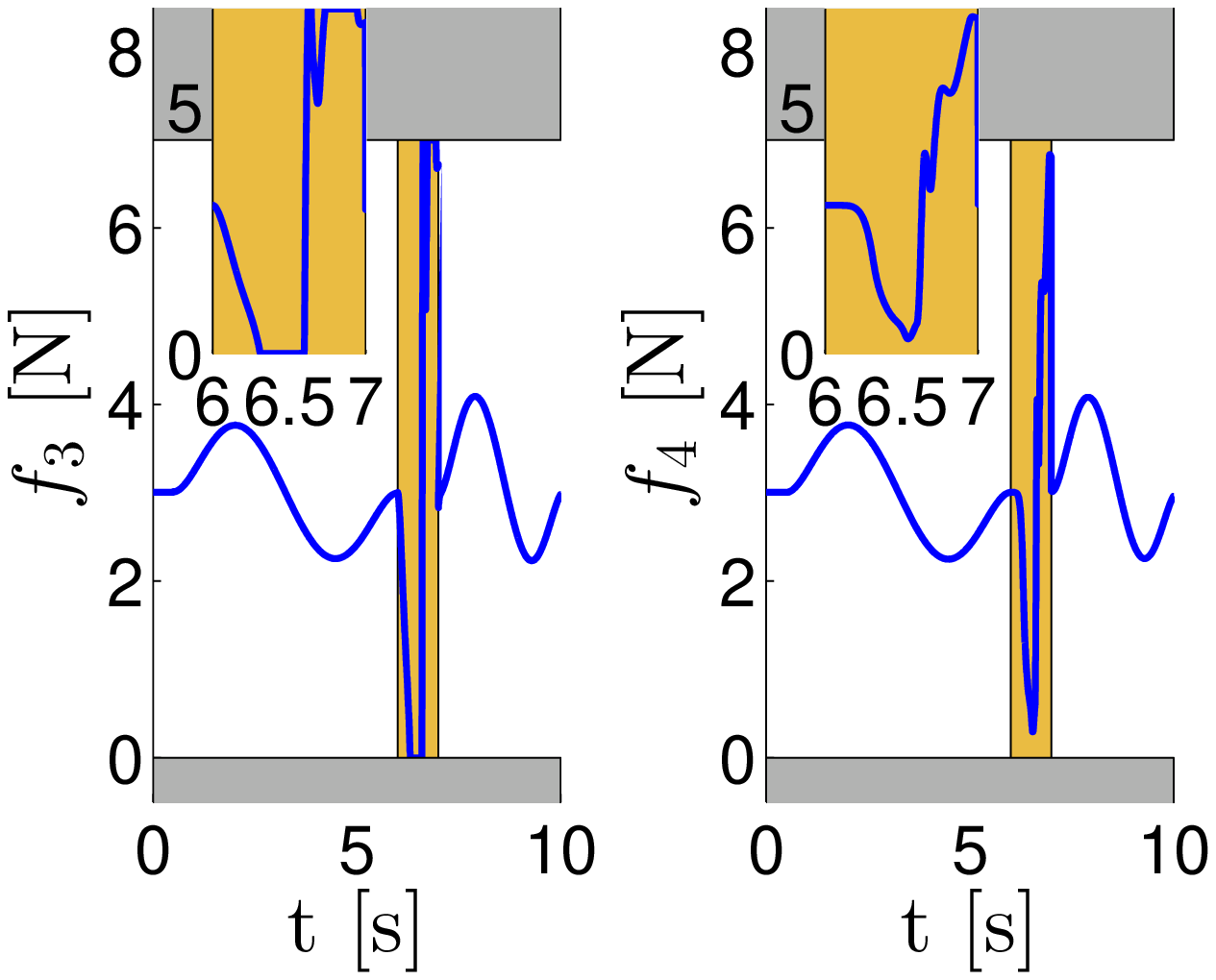}}
\put(-78,75){\parbox{\columnwidth}{${\nwarrow}$}}
\put(-68,68){\parbox{\columnwidth}{sat}}
\put(-94,25){\parbox{\columnwidth}{${\searrow}$}}
\put(-104,33){\parbox{\columnwidth}{sat}}

\caption{
Complex trajectory tracking.
Black lines: Developed Null-space (NS). Blue dashed lines: Benchmark.
(\ref{PsiComp}) Attitude error given by (\ref{error_function_A}). 
(\ref{normgon}) Angular velocity error, $\lVert\mathbf{e}_{\omega}\rVert$.
(\ref{normpos}) Position error, $\lVert\mathbf{e}_{x}\rVert$.
(\ref{trajU}) Trajectory response comparison with $f_{p}{=}0$.
(\ref{thrR11},\ref{thrR12}) Thrusts (Developed NS).
(\ref{thrM11},\ref{thrM12}) Thrusts (Benchmark).
Orange shaded intervals: Attitude Mode.}
\label{Aggressive}
\end{figure}

Simulation results of the maneuver are illustrated in Fig. \ref{Aggressive}.
The time during which the Attitude Mode is employed is illustrated by the orange shaded intervals.
The ability of the developed flight modes in achieving precise trajectory tracking during the Position Mode is apparent.
Examining Figs. \ref{PsiComp}, \ref{normgon}, \ref{normpos}, it is observed that the attitude error $\Psi$, the angular velocity error $\lVert\mathbf{e}_{\omega}\rVert$, and the position error $\lVert\mathbf{e}_{x}\rVert$, only increase during the attitude portion of the maneuver (see Fig. \ref{PsiComp}, \ref{normgon}, \ref{normpos}, $6{\leq}t{<}7$).
This is due to multiple reasons.
First the quadrotor is underactuated and thus it can not track simultaneously both the desired attitude and desired position, secondly the motors saturate multiple times and finally the $360^{o}$ flip is quite aggressive.
The underactuated nature of the quadrotor is the cause of the large position error during the Attitude Mode, while motor saturation is the cause for the increase in attitude and angular velocity error.
However when the Null-space solution is used, we have almost perfect tracking of the attitude/angular velocity objectives and less position error in comparison to the Benchmark solution.

Specifically the Null-space solution, during the $360^{o}$ flip maneuver, demonstrates an increase only in the position tracking error, $\lVert\mathbf{e}_{x}\rVert{<}1.5529$ [m] (see Fig. \ref{normpos}, $6{\leq}t{<}7$).
The attitude error remains below $\Psi{\leq}2.8564{\cdot}10^{-9}$ ($2.5708{\cdot}10^{-7}$ [deg] with respect to an axis-angle rotation) meaning that the attitude is tracked exactly, (see magnified insert in Fig. \ref{PsiComp}, thick black line), while ${}^{b}\boldsymbol{\omega}_{d}(t)$ is tracked faithfully, with $\lVert\mathbf{e}_{\omega}\rVert{\leq}0.0028 rad/s$ (see magnified insert in Fig. \ref{normgon}, thick black line).
During the same time period ($6{\leq}{t}{<}7$) the benchmark solution results in considerably higher tracking errors compared to the developed one.
In particular, the attitude error of the Benchmark solution remains below $\Psi{<}7.3682{\cdot}10^{-4}$ (0.0663 [deg] wrt., an axis-angle rotation) denoting an error $2.5795{\cdot}10^{5}$ times worse compared to the developed one.
It is clear that the developed Null-space solution outperforms by far the Benchmark one.
The same holds for the angular velocity error where the Benchmark with $\lVert\mathbf{e}_{\omega}\rVert{\leq}1.5347 rad/s$ (see Fig. \ref{PsiComp},\ref{normgon}, thin blue line) exhibits an error more than $543$ times worse.
During the $360^{o}$ flip, the Benchmark position error is $\lVert\mathbf{e}_{x}\rVert{<}1.7181$ [m] (see dashed line on Fig. \ref{normpos}) signifying an error 1.1064 times worse.
Again the developed Null-space solution performs better.

The above results are attributed to the Null-space solution ability to produce thrusts that do not saturate the actuators, and still serve the attitude control objective (see Fig. \ref{PsiComp},\ref{thrR11},\ref{thrR12}).
Indeed, the lowest registered thrust is 0.4769 [N] while the largest equals to 6.7601 [N] and complies with the actuator constraints (see Fig. \ref{thrR11},\ref{thrR12}).
In contrast to this, the Benchmark solution during the Attitude Mode, is prone to thruster saturation (see 'sat' in Fig. \ref{thrM11}, \ref{thrM12}). 
However this implies that the developed attitude controller, (\ref{att_contr}), utilized in the Benchmark solution is robust to thruster saturation.

During the attitude maneuver, the developed solution is also able to ''track'' a desired position command, in the margins allowed by the actuator constraints, through the null-space projection of $f_{p}$ (see (\ref{xi})).
To comprehend better the effects of $f_{p}$, the same simulation was performed, but this time with $f_{p}{=}0$.
The results can be seen in Fig. \ref{trajU}, where the black solid lines correspond to tracking error with active $f_{p}$, while the blue dashed lines correspond to $f_{p}{=}0$.
By comparing the responses in Fig. \ref{trajU}, it is clear that $f_{p}$ achieves its goal satisfactory well.
With $f_{p}$ absent, the position deviation exceeds 0.6 [m] in the $x_{1}$ direction and 2.535 [m] in the $x_{3}$ direction (see Fig. (\ref{trajU}), $6{\leq}{t}{<}7$).
In contrast to this, with $f_{p}$ present, the position deviation has a mean value of close to zero ($\mu_{e_{x_1}}{=}-0.1274$) in the $x_{1}$ direction and remains below 1.55 [m] in the $x_{3}$ direction (see Fig. (\ref{trajU}), $6{\leq}{t}{<}7$).
Note that to improve legibility, $e_{x_{2}}$ in Fig. \ref{trajU} has been omitted because this maneuver results to a rotation purely around $\mathbf{e}_{1}$, and no translation in the direction of $\mathbf{E}_{2}$ exists; thus $e_{x_{2}}{=}0m$.
It is emphasized that the position tracking objective is achieved as a secondary task in the margins allowed by actuator constraints.
As a result the developed solution is intended for short durations of time.
Nevertheless the ability of the null-space solution, to briefly track a desired position command while complying to actuator constraints, without interfering with the attitude control objective is verified.

Note that the guarantees produced by the stability proofs hold throughout the maneuver, since the thrusts produced by the developed solution adhere to motor constraints.
Through the simulations, the developed solution showcased results of increased precision that could be deemed redundant, nevertheless the results are supplemented with guarantees on the system performance.
Finally since the generated thrusts, during the attitude maneuver are not negative, they are realizable by standard outrunner motors and thus by the majority of quadrotors produced by the industry, an important contribution.

\section{Conclusion and Future Work}

The  quadrotor task of negotiating aggressive attitude maneuvers while adhering to motor constraints was addressed here.
An attitude control framework was developed, comprised by a thrust allocation strategy and a specially designed geometric attitude tracking controller, allowing the quadrotor to achieve aggressive attitude maneuvers, while complying to actuator constraints, and simultaneously staying ''close'' to a desired position command in a computationally inexpensive way.
This is a novel contribution resulting in thrusts realizable by available quadrotors during aggressive attitude maneuvers, and enhanced performance guaranteed by valid stability proofs.
Additionally, it was shown that the developed controller can be combined with a collective thrust expression in producing a position/yaw tracking controller.
Through rigorous stability proofs, both the position and attitude frameworks were shown to have desirable closed loop properties that are almost global.
This established a quadrotor control solution, allowing the vehicle to negotiate aggressive maneuvers position/attitude on SE(3).
Simulations illustrated and validated the effectiveness and capabilities of the developed solution.

Our future work will include experimental trials on the feasibility of the proposed strategy and controller.


%


%

%
\appendices
\section{\label{appA}}
The attitude tracking errors associated with the attitude error function studied in \cite{err_fun}, and related   properties are given next.

\textbf{Proposition \arabic{Prop1}.}
In regards to $(\ref{error_function_A})$-$(\ref{ang_vel_error})$, the attitude error vector, $(\ref{att_error_A})$, is well defined in (\ref{L_2}).
Thus for a tracking command $(\mathbf{R}_{d},{}^{b}\boldsymbol{\omega}_{d})$ and current state $(\mathbf{R},{}^{b}\boldsymbol{\omega})$,
\begin{enumerate}[(i)]
\setcounter{enumi}{\value{numberlatin}}
\item $\Psi$ is locally positive-definite about $\mathbf{R}=\mathbf{R}_{d}$.
\item In (\ref{L_2}) the left-trivialized derivative of $\Psi$ is given by,
\begin{IEEEeqnarray}{rCl}
\text{T}^{*}_{I}\text{L}_{R}(\mathbf{D}_{R}\Psi(\mathbf{R},\mathbf{R}_{d}))=\mathbf{e}_{R}
\label{left_triv_der_A}
\end{IEEEeqnarray}
\item The critical points of $\Psi$, where $\mathbf{e}_{R}=0$, are $\{\mathbf{R}_{d}\}\cap\{\mathbf{R}_{d}\text{exp}(\pi S(\mathbf{s})),\mathbf{s}\in\text{S}^{2}\}$ and there exists only one critical point $\{\mathbf{R}_{d}\}$ in (\ref{L_2}).
\item $\Psi$ is locally quadratic in (\ref{L_2}), since
\begin{IEEEeqnarray}{rCl}
\lVert\mathbf{e}_{R}(\mathbf{R},\mathbf{R}_{d})\rVert^{2}\leq &\Psi(\mathbf{R},\mathbf{R}_{d})&\leq 2\lVert\mathbf{e}_{R}(\mathbf{R},\mathbf{R}_{d})\rVert^{2}
\label{quadr_Psi}
\end{IEEEeqnarray}
\end{enumerate}
\textbf{Proof of Proposition \arabic{Prop1}.} 
See \cite{err_fun} for statements (\rom{1})-(\rom{4}).

The associated attitude error dynamics of (\ref{error_function_A})-(\ref{ang_vel_error}) to be used in the control design are given next.

\textbf{Proposition \arabic{Prop2}.} The error dynamics of $\{(\ref{error_function_A}),(\ref{att_error_A})\}$, satisfy:
\begin{IEEEeqnarray}{rCl}
\dot{\Psi}(\mathbf{R},\mathbf{R}_{d})&=&\mathbf{e}_{R}^{T}\mathbf{e}_{\omega}\label{dot_Psi_A}\\
\dot{\mathbf{e}}_{R}&=&\mathbf{E}(\mathbf{R},\mathbf{R}_{d})\mathbf{e}_{\omega}\label{dot_Att_Error_A}\\
\mathbf{E}(\mathbf{R},\mathbf{R}_{d})&=&\frac{\{ tr [ \mathbf{R}^{T}\mathbf{R}_{d} ] \mathbf{I}-\mathbf{R}^{T}\mathbf{R}_{d}+2\mathbf{e}_{R}\mathbf{e}_{R}^{T} \}}{2\sqrt{1+tr[\mathbf{R}^{T}_{d}\mathbf{R}]}}\\
\lVert\dot{\mathbf{e}}_{R}\rVert&\leq&\frac{1}{2}\lVert\mathbf{e}_{\omega}\rVert,\lVert\mathbf{E}(\mathbf{R},\mathbf{R}_{d})\rVert=\frac{1}{2}\label{erDot_norm_A}
\end{IEEEeqnarray}

The time derivative of (\ref{ang_vel_error}) is given by,
\begin{IEEEeqnarray}{rCl}
\dot{\mathbf{e}}_{\omega}&=&{}^{b}\dot{\boldsymbol{\omega}}+\mathbf{a}_{d}\IEEEnonumber\\
&=&\mathbf{J}^{-1}\left({}^{b}\mathbf{u}-{}^{b}\boldsymbol{\omega}\times\mathbf{J}{}^{b}\boldsymbol{\omega}\right)+\mathbf{a}_{d}\IEEEyesnumber\IEEEeqnarraynumspace
\label{att_error_dyn}\\
\mathbf{a}_{d}&=&S({}^{b}{\boldsymbol{\omega}})\mathbf{R}^{T}\mathbf{R}_{d}{}^{b}{\boldsymbol{\omega}}_{d}-\mathbf{R}^{T}\mathbf{R}_{d}{}^{b}{\dot{\boldsymbol{\omega}}}_{d}
\label{E_ad}
\end{IEEEeqnarray}
\textbf{Proof of Proposition \arabic{Prop2}.}
See \cite{err_fun}.

\end{document}